\documentclass[amsmath,pre,superscriptaddress,preprint]{revtex4}
\usepackage{graphicx,graphics}
\usepackage{mathrsfs}
\usepackage{color}

\begin{document}  
 
\title{Dynamic Computation of Network Statistics \\ via \\ Updating Schema}
\author{Jie Sun}
\email{sunj@clarkson.edu} 
\affiliation{Department of Mathematics \& Computer Science, Clarkson University, Potsdam, NY 13699-5815, USA}
\author{James P.~Bagrow}
\email{j.bagrow@neu.edu}
\affiliation{Department of Physics, Clarkson University, Potsdam, NY 13699-5820, USA}
\affiliation{Center for Complex Networks Research and Department of Physics, Northeastern University, Boston, MA 02115, USA}
\author{Erik M.~Bollt}
\email{bolltem@clarkson.edu}
\affiliation{Department of Mathematics \& Computer Science, Clarkson University, Potsdam, NY 13699-5815, USA}
\affiliation{Department of Physics, Clarkson University, Potsdam, NY 13699-5820, USA}
\author{Joesph D.~Skufca}
\email{jskufca@clarkson.edu}
\affiliation{Department of Mathematics \& Computer Science, Clarkson University, Potsdam, NY 13699-5815, USA}

\begin{abstract}
In this paper we derive an updating scheme for calculating some important network statistics such as degree, clustering coefficient, etc., aiming at reduce the amount of computation needed to track the evolving behavior of large networks; and more importantly, to provide efficient methods for potential use of modeling the evolution of networks. Using the updating scheme, the network statistics can be computed and updated easily and much faster than re-calculating each time for large evolving networks. The update formula can also be used to determine which edge/node will lead to the extremal change of network statistics, providing a way of predicting or designing evolution rule of networks.
\end{abstract}

\maketitle

\section{Introduction}
Complex networks are useful tools for modeling complicated real life objects and their interactions. Examples include computer networks, social networks, biological networks, etc. \cite{WATTS_NATURE98}\cite{BARABASI_SCIENCE99}\cite{ALBERT_REV02}\cite{SATORRAS_PRL01}\cite{VAZQUEZ_PRE02}\cite{ShiZhou_PRE06}. Different from traditional graph theory approach which emphasize on micro-state quantity of each node in the network, recently developed statistical methods \cite{ALBERT_REV02} allow us to analyze large networks by summarizing several important statistics out of a massive amount of information carried by the network itself.  These statistics include degree (number of connections each node has), clustering coefficient \cite{WATTS_NATURE98}, assortativity coefficient \cite{NEWMAN_PRL02}, modularity measure \cite{NEWMAN_PNAS06}, etc. Fast algorithms \cite{NetworkX}\cite{Pajek} have been developed to compute these statistics for any given network, either represented by adjacency matrix or edge list \cite{CORMEN_BOOK}. 

However, for any evolving network, to measure the corresponding evolution of network statistics, the computation based on static network structure has to be done for the network at each time step, resulting in an impractical task even each single computation is fast. A missing part in the study of evolving network is a development of a dynamic algorithm which updates, rather than re-compute the statistics. 

In this paper we present a dynamical algorithm based on the knowledge of existing network structure and the changes to the network. We will consider adjacency matrix as the default structure of representing a network \footnote{The formula can be adopted for edge list easily, while for pedagogical reasons we present adjacency matrix format.}. The results hold very similarly if one uses edge-list instead.

The rest of the paper is organized as follows. In Section II we review the definition of some network statistics and introduce notation that will be used in the paper for these statistics. In Section III we derive update formula for network statistics upon the change of network structure and compare the computational complexity to the use of regular methods. In Section IV we show examples of application using updating scheme. In Section V we discuss the main results of the paper and give some overview of potential future research.

\section{Definition and Notation}
A mathematical representation of a network is a graph $G=(V,E)$ where $V=\{1,2,...,N\}$ is the vertex set and $E=\{(i,j)\,|\, \text{$i$ and $j$ are connected}\}$ is the edge set. Note that for undirected graphs, if $(i,j)\in E$ then so is $(j,i)$ \footnote{For undirected graphs we could have eliminated the element $(j,i)$ in $E$ if $(i,j)$ has been put in $E$, but including both of them allows easier generalization to directed graphs.}. In this work, we limit ourselves to undirected, unweighted networks;  their graphs possess a symmetric, binary {\it adjacency matrix} $A$:
\begin{equation}
a_{ij} = 
\begin{cases} 1, & \text{if $(i,j)\in E$;}\\
0, & \text{otherwise.}
\end{cases}
\end{equation}
With $M$ as the total number of edges in $G$, then
\begin{equation}\label{edges}
	M \equiv \frac{1}{2}|E|= \frac{1}{2}||A||_F^2 = \frac{1}{2}\sum_{i,j}{a_{ij}}.
\end{equation}
Here $|.|$ is the cardinality of a set.

Define the neighborhood $N(i)$ of node $i$ as the set of vertices that are adjacent to $i$, i.e.:
\begin{equation}\label{nbs}
	N(i) \equiv \left\{j|(i,j)\in E\right\} = \left\{j|a_{ij}=1\right\}.
\end{equation}
Likewise, define the \emph{shared} neighborhood $N_{ij}$ of nodes $i$ and $j$ as:
\begin{equation}\label{nbs}
	N_{ij} \equiv N(i) \cap N(j).
\end{equation}
The {\it degree}  $k_{i}$ of node $i$ is the number of nodes it connects to:
\begin{equation}\label{deg}
	k_{i} \equiv  \left| N(i) \right| = \sum_j a_{ij} = \sum_j a_{ji},
\end{equation}
since we limit ourselves to undirected networks.

The {\it clustering coefficient} of node $i$ is defined by \cite{WATTS_NATURE98}:
\begin{equation}\label{clcoef}
	C_{i} \equiv
	\begin{cases} \frac{2\triangle_{i}}{k_{i}(k_{i}-1)}, & \text{if $k_{i}\geq 2$};\\
		0, & \text{otherwise,}
	\end{cases}
\end{equation}
where $\triangle_{i}$ is the number of triangles that contain $i$. Then the {\it average clustering coefficient} \footnote{There is an alternative definition of the average clustering coefficient of a network in \cite{NEWMAN_SIAM03}, and the formula for updating this alternative $C$ can be derived easily based on updating the number of triangles and triples in the network.} of the whole network is simply the average of all $C_{i}$'s:
\begin{equation}\label{netclcoef}
	C \equiv \frac{1}{N}\sum_{i}{C_{i}}.
\end{equation}

Another interesting quantity is the {\it assortativity coefficient} $r$ \cite{NEWMAN_PRL02} of a network:
\begin{eqnarray}\label{asscoef}
	r &\equiv&  \frac{8M\sum_{(i,j)\in E}{k_{i}k_{j}} - \left[\sum_{(i,j)\in E}{\left(k_{i}+k_{j}\right)}\right]^2}{4M\sum_{(i,j)\in E}{\left(k_{i}^2+k_{j}^2\right)} - \left[\sum_{(i,j)\in E}{\left(k_{i}+k_{j}\right)}\right]^2}  \nonumber\\
	&=& \frac{8Mu-v^2}{4Mw-v^2}, 
\end{eqnarray}	
where
\begin{eqnarray}\label{defuvw}
	u &\equiv& \sum_{(i,j)\in E}{k_{i}k_{j}}, \quad	\\
	v &\equiv& \sum_{(i,j)\in E}{\left(k_{i}+k_{j}\right)}, \quad  \\ 
	w &\equiv& \sum_{(i,j)\in E}{\left(k_{i}^2+k_{j}^2\right)}.
\end{eqnarray}

{\it Modularity} $Q$ \cite{NEWMAN_PNAS06} is a quantity which measures the quality of a community partition, typically defined as:
\begin{align}\label{modularity}
	Q &\equiv  \frac{1}{2M}\sum_{i,j}\left(a_{ij} - \frac{k_{i}k_{j}}{2M}\right)\delta(g_i,g_j) \nonumber\\
\	      &= \frac{1}{2M}\left[S_A-\frac{1}{2M}S_P\right],
\end{align}
where $\delta(g_i,g_j)=1$ if nodes $i$ and $j$ are in the same group and zero otherwise, and
\begin{equation}\label{modularityFH}
	S_A \equiv \sum_{i,j}a_{ij}\delta(g_i,g_j), \quad 	S_P \equiv \sum_{i,j}{k_{i}k_j \delta(g_i,g_j)}.
\end{equation}

\section{Updating Schemes for Statistics upon Local Information}
\subsection{Adding an edge between existing nodes}
Suppose $a_{pq} = 0$ ($p\neq q$ and $p$, $q$ are not connected), we analyze the impact of connecting $p$ and $q$ on the various statistics of the network. The goal is to derive computations that are as inexpensive as possible.  We use $\,\widetilde{\cdot}\,$ to represent updated statistics:
\begin{gather}
	\widetilde{E} = E\cup \{(p,q),(q,p)\},\\
	\widetilde{M}  =  M + \Delta^{+}{M} = M+1,\label{eqn:numEdges_Update}
\end{gather}
and
\begin{equation}
\widetilde{a}_{ij} = a_{ij} + \Delta^{+}{a_{ij}} = a_{ij} + \delta_{ip} \delta_{jq} +  \delta_{iq} \delta_{jp},
\end{equation}
where we use {\it update delta} $\Delta^{+}$ to represent the change for statistics upon adding an edge to the existing network, and in the following $\Delta^{-}$ will be used to denote change for statistics upon deleting an existing edge. We will not explicitly specify which edge to add or delete in the update delta notation when there is no confusion.

Based on the above formulas, we can derive schemes for efficiently updating network statistics.

{\bf Degree}

The change in degree for node $i$ is simply:
\begin{equation}\label{newdeg}
	\widetilde{k}_{i} = k_i + \Delta^{+}{k_i} = k_i + \delta_{ip} + \delta_{iq},
\end{equation}
where
\begin{equation}
	\Delta^{+}{k_i} = \delta_{ip} + \delta_{iq}.
\end{equation}
The above formula indicates that the degree changes only for vertex $p$ and $q$, so that if one keeps a list of the degree of all vertices of the network, each update takes only $2$ operations when a new edge is added.

{\bf Clustering Coefficient}

To compute the new clustering coefficient of each node, and thus the whole network, we need the updated number of triangles at node $i$:
\begin{equation}
	\widetilde{\triangle}_{i}  = 
	\begin{cases}
	\triangle_{i}, & \text{if $i\notin\left\{p,q\right\}\cup N_{pq}$};\\
	\triangle_{i} + 1, &\text{if $i\in N_{pq}$};\\
	\triangle_{i} + \left| N_{pq} \right|, & \text{if $i\in\{p,q\}$}.
	\end{cases}
\end{equation}
Combining this with Eq.~\eqref{newdeg} and $\triangle_{i}=\frac{1}{2}C_{i}k_{i}(k_{i}-1)$, from Eq.~\eqref{clcoef},  we have:
\begin{equation}
	\widetilde{C}_{i} =
	\begin{cases}
		C_{i}, & \text{if $i\notin\{p,q\}\cup N_{pq}$};\\
		C_{i} + \frac{2}{k_{i}(k_{i}-1)}, & \text{if $i \in N_{pq}$};\\
		\frac{k_{i}-1}{k_{i}+1}C_{i}+\frac{2 \left| N_{pq}\right|}{k_i\left(k_i+1\right)}, & \text{if $i\in\{p,q\}$}.
	\end{cases} \label{eqn:newClustering}
\end{equation}
Note that whenever the denominator of a fraction is zero, we define the fraction to be zero, in Eq.~\eqref{eqn:newClustering} and throughout.  This maintains the consistency that $C_i = 0$ if $k_i < 2$. 
Finally, the average clustering coefficient $C$ becomes:
\begin{equation}
	\widetilde{C}
	  = C + \Delta^{+}{C}
	= C + \frac{2}{N}\left[ \sum_{i\in N_{pq}}{\frac{1}{k_{i}(k_{i}-1)}} + \sum_{i\in\{p,q\}}{\left(\frac{\left| N_{pq}\right|}{k_{i}(k_{i}+1)} - \frac{C_{i}}{k_{i}+1}\right)}  \right]
\end{equation}
where
\begin{equation}
	\Delta^{+}{C} = \frac{2}{N}\left[ \sum_{i\in N_{pq}}{\frac{1}{k_{i}(k_{i}-1)}} + \sum_{i\in\{p,q\}}{\left(\frac{\left| N_{pq}\right|}{k_{i}(k_{i}+1)} - \frac{C_{i}}{k_{i}+1}\right)}  \right].
\end{equation}

Note that to update the average clustering coefficient, we need to keep the clustering coefficient for each node in order to apply the update formula, which implies an $O(N)$ storage complexity.

{\bf Assortativity Coefficient}

To compute $\widetilde{r}$, we need $\widetilde{u}$, $\widetilde{v}$, and $\widetilde{w}$. The update formula for $u$ is:
\begin{align}
	\widetilde{u} &= \sum_{(i,j)\in \widetilde{E}}{\widetilde{k}_{i}\widetilde{k}_{j}} = \sum_{(i,j)\in E}{\widetilde{k}_{i}\widetilde{k}_{j}} + 2(k_{p}+1)(k_{q}+1) \nonumber\\
         &= \sum_{(i,j)\in \widehat{E}}{ k_{i}k_{j}} + 2\sum_{i\in N(p)}{k_{i}(k_{p}+1)} \nonumber\\
         &\mbox{  }+2\sum_{i\in N(q)}{k_{i}(k_{q}+1)} + 2(k_{p}+1)(k_{q}+1) \nonumber\\
         &= u + 2\left(\sum_{i\in N(p)}{k_{i}} + \sum_{i\in N(q)}{k_{i}}\right) + 2(k_{p}+1)(k_{q}+1)\nonumber\\
         &= u + \Delta^{+}{u}.
\end{align}
Here $\widehat{E} = E \setminus \{(p,q),(q,p)\}$ is the edge set that contains all edges in $E$ but $(p,q)$ and $(q,p)$ and 
\begin{equation}
	\Delta^{+}{u} =2\left(\sum_{i\in N(p)}{k_{i}} + \sum_{i\in N(q)}{k_{i}}\right) + 2(k_{p}+1)(k_{q}+1).
\end{equation}

Similarly, we can obtain update formula for $v$ and $w$:
\begin{align}
	\widetilde{v} &= \sum_{(i,j)\in \widetilde{E}}{(\widetilde{k}_{i} + \widetilde{k}_{j})}\nonumber\\
         &= v + 4(k_{p}+k_{q}+1) = v + \Delta^{+}{v},
\end{align}
where
\begin{equation}
	\Delta^{+}{v} = 4(k_{p}+k_{q}+1).
\end{equation}
For $w$ we have:
\begin{align}
	\widetilde{w} &= \sum_{(i,j)\in \widetilde{E}}{(\widetilde{k}_{i}^2 + \widetilde{k}_{j}^2)} \nonumber\\
         &=  w + \Delta^{+}{w},
\end{align}
where
\begin{equation}
	\Delta^{+}{w} = 6\left[k_{p}(k_{p}+1)+k_{q}(k_{q}+1)\right] + 4.
\end{equation}

Finally, the new assortativity coefficient can be updated using:
\begin{eqnarray}
	\widetilde{r} &=& r + \Delta^{+}{r} \nonumber\\ 
	  &=& \frac{8\widetilde{M}\widetilde{u}-\widetilde{v}^2}{4\widetilde{M}\widetilde{w}-\widetilde{v}^2} \nonumber\\
	&=& \frac{8\left(M+1\right)\left(u+\Delta^{+}{u}\right) - \left(v+\Delta^{+}{v}\right)^2}{4\left(M+1\right)\left(w+\Delta^{+}{w}\right) - \left(v+\Delta^{+}{v}\right)^2}.
\end{eqnarray}

{\bf Modularity}

For modularity, we assume that after connecting the nodes $p$ and $q$, the partitions $g_i$  do not change for any node $i$. Then the new modularity measure will be:
\begin{equation}
	\widetilde{Q} = \frac{1}{2\widetilde{M}}\left[\widetilde{S}_A-\frac{1}{2\widetilde{M}}\widetilde{S}_P\right].
	\label{eqn:updatedAddedQ}
\end{equation}
We already have $\widetilde{M} = M+1$, we now derive updating formulas for $S_A$ and $S_P$. By Eq.~(\ref{modularityFH}), we have:
\begin{align}
	\widetilde{S}_A &= S_A + \Delta^{+}S_A \nonumber\\
	&= \sum_{i,j}{\widetilde{a}_{ij}\delta(g_i,g_j)} \nonumber\\
	&=\sum_{i,j} \left( a_{ij} +  \delta_{ip} \delta_{jq} +  \delta_{iq} \delta_{jp} \right)\delta(g_i,g_j)   \nonumber\\
	&= S_A + 2 \delta(g_p,g_q)
\label{eqn:adj_sumUpdate}
\end{align}
where 
$\Delta^{+}S_A = 2 \delta(g_p,g_q)$;

and
\begin{align}
	\widetilde{S_P} &= S_P + \Delta^{+}S_P \nonumber\\
	  &= \sum_{i,j}{\widetilde{k}_{i}\widetilde{k}_j}\delta(g_i,g_j) \nonumber\\
	&= \sum_{i,j}\left( k_i + \delta_{ip} + \delta_{iq}\right)\left( k_j + \delta_{jp} + \delta_{jq}\right) \delta(g_i,g_j)\nonumber\\
				    &= S_P + 2\sum_i k_i \big[\delta(g_i,g_p)+\delta(g_i,g_q)\big] +2 \big[\delta(g_p,g_q)+1 \big].
	\label{eqn:null_modelTermUpdateAlmost}
\end{align}

However, computing the sum in Eq.~\eqref{eqn:null_modelTermUpdateAlmost} for every update is expensive.  To avoid this, define the following auxiliary statistics:
\begin{equation}
	K_g \equiv \sum_i k_i \delta(g_i,g)
\end{equation}
with updating scheme
\begin{eqnarray}
	\widetilde{K}_g &=& K_g + \Delta^{+}{K_g} \nonumber\\
		&=& K_g + \delta(g_p,g) + \delta(g_q,g)
\end{eqnarray}
giving
\begin{equation}
	\widetilde{S}_P  = S_P + \Delta^{+}S_P = S_P + 2\left(K_{g_p} + K_{g_q}\right) +2 \big[\delta(g_p,g_q)+1 \big]
	\label{eqn:null_modelTermUpdate}
\end{equation}
where
$\Delta^{+}S_P = 2\left(K_{g_p} + K_{g_q}\right) +2 \big[\delta(g_p,g_q)+1 \big].$

Finally, combining \eqref{eqn:adj_sumUpdate} and \eqref{eqn:null_modelTermUpdate} with \eqref{eqn:updatedAddedQ} gives the updating scheme for $Q$:
\begin{eqnarray}\label{updateQ}
	\widetilde{Q} &&= Q + \Delta^{+}{Q} \nonumber\\
		&&= \frac{1}{2(M+1)} \Big[ S_A + 2\delta(g_p,g_q) - \frac{1}{2(M+1)}\Big(S_p + 2[K_{g_p}+K_{g_q}]+2[\delta(g_p,g-q)+1]\Big) \Big]. \nonumber\\
\end{eqnarray}

\begin{figure}[ht]
\includegraphics*[width=0.55\textwidth]{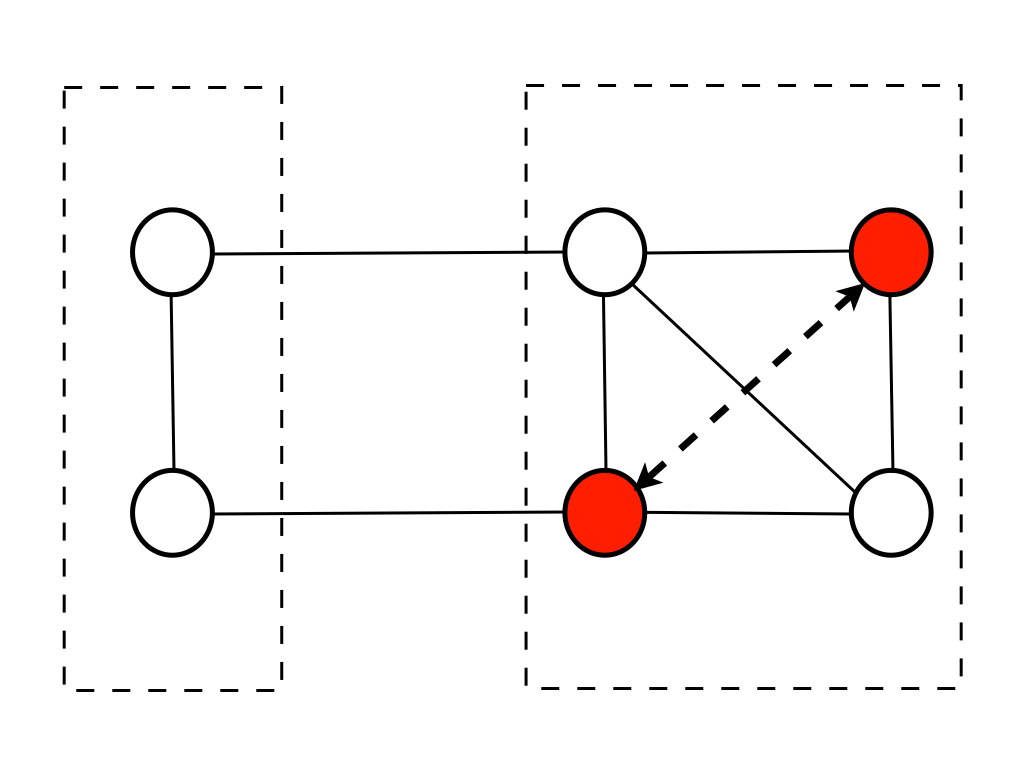}
\caption{(Color online) An example that the modularity actually decreases when a new edge is added to vertices within the same group. The dashed oval boxes indicate the preexisting partition of the graph into two groups. Solid lines are the edges in the original graph. Before adding the new edge (dashes arrow line), the modularity is $0.125$. After a new edge is added between two vertices in the same group (solid circles) the updated modularity becomes $0.1235$.}
\label{graphpartitions}
\end{figure}

From Eq.~(\ref{updateQ}) one is able to predict whether the modularity measure $Q$ increases or decreases with the knowledge of existing partition of the graph as well as the edge to be added. For example, if there is a preexisting partition of the graph into two groups, then if a new edge is added in between the two groups, then $\Delta^{+}Q<0$, i.e., the modularity is to decrease. On the other hand, if a new edge is added to vertices belonging to the same group, then the modularity increases if the edge is added to the group with smaller total degree; However, adding an edge within a group does not necessarily increase $Q$ if the edge is added into a group with larger total degree, see Fig.~\ref{graphpartitions} as an example.

\subsection{Connecting a New Node}
The operation of adding an edge to a new node can be decomposed into two successive operations: first, introduce an isolated node that connects to nothing in the network; then add an edge between this node a previously existing node. We can use the previous results for the second step and need only focus on the first, i.e. adding an empty node to a network.

Since no new edge is introduced, it's easy to obtain the following updating relations:
\begin{equation}
	\widetilde{N} = N+1, \quad \widetilde{M} = M, \quad \widetilde{E} = E,
\end{equation}
and
\begin{equation}
	\widetilde{a}_{ij} =
	\begin{cases}
		a_{ij}, & \text{if $i\neq{N+1}$ and $j\neq{N+1}$};\\
		0, & \text{otherwise}.
	\end{cases}
\end{equation}
Then for other statistics, we have:
\begin{align}
	\widetilde{k}_{i} &= k_{i},\quad i \neq N+1;\nonumber\\
	\widetilde{k}_{N+1} &= 0;
\end{align}
and
\begin{align}
	\widetilde{C}_{i} &= C_{i},\quad i \neq N+1;\nonumber\\
	\widetilde{C}_{N+1} &= 0,
\end{align}
so that
\begin{equation}
	\widetilde{C} = \frac{1}{\widetilde{N}}\sum_{i}{\widetilde{C}_{i}} = \frac{1}{N+1}\sum_{i}{C_{i}} = \frac{N}{N+1}C.
\end{equation}
Similarly, $\widetilde{r}=r$ since $\widetilde{u} = u$, $\widetilde{v} = v$, and $\widetilde{w} = w$; and $\widetilde{Q}=Q$ since $\widetilde{F} = F$, and $\widetilde{H} = H$.

\subsection{Deleting an Existing Edge}
Now we investigate how network statistics changes when we delete an existing edge in the network. Suppose $a_{pq} = 1$ ($p\neq q$ and $p,q$ are connected), and we delete this edge, $(p,q)\cup (q,p)$, from our edge set $E$. Using $\widehat{A}$ to represent the updated adjacency matrix, and similarly for other statistics. Then we immediately have: 
\begin{gather}
	\widehat{E} = E \setminus \{(p,q),(q,p)\},\\
	\widehat{M} = M-1,
\end{gather}
and
\begin{equation}
	\widehat{a}_{ij} = a_{ij} + \Delta^{-}{a_{ij}} = a_{ij} - \delta_{ip} \delta_{jq} - \delta_{iq}\delta_{jp.}
\end{equation}

{\bf Degree}

The change in degree for node $i$ is:
\begin{equation}
	\widehat{k}_{i} = k_i + \Delta^{-}{k_i}  = k_i - \delta_{ip} - \delta_{iq}
\end{equation}
where
\begin{equation}
	\Delta^{-}{k_i}  = - \delta_{ip} - \delta_{iq}.
\end{equation}

{\bf Clustering Coefficient}

For the new clustering coefficient, we first obtain the formula for updating the number of triangles containing node $i$:
\begin{equation}
	\widehat{\triangle}_{i} =
	\begin{cases}
	 \triangle_{i},& \text{if $i\notin\{p,q\}\cup N_{pq}$};\\
	\triangle_{i} - 1,& \text{if $i\in N_{pq}$};\\
	\triangle_{i} - |N_{pq}|,& \text{if $i\in\{p,q\}$}.
	\end{cases}
\end{equation}
Then we obtain the formula for updating $C_{i}$:
\begin{equation}
	\widehat{C_{i}} = 
	\begin{cases}
	C_{i},& \text{if $i\notin\{p,q\}\cup N_{pq}$};\\
	C_{i} - \frac{2}{k_{i}(k_{i}-1)},& \text{if $i\in N_{pq}$};\\
	\frac{k_{i}}{k_{i}-2}C_{i}-\frac{2 \left|N_{pq}\right|}{(k_{i}-1)(k_{i}-2)}, & \text{if $i\in\{p,q\}$}.
	\end{cases}	
\end{equation}
The average clustering coefficient $C$ is updated by:
\begin{eqnarray}
	\widehat{C} &=& C + \Delta^{-}{C} \nonumber\\
		&=& \frac{1}{N}\sum_{i}{\widehat{C_{i}}} \nonumber\\ 
	&=& C - \frac{2}{N}\left[ \sum_{i\in{N_{pq}}}{\frac{1}{k_{i}(k_{i}-1)}} + \sum_{i\in\{p,q\}}{\left(\frac{|N_{pq}|}{(k_{i}-1)(k_{i}-2)} - \frac{C_{i}}{k_{i}-2}\right)}  \right],
\end{eqnarray}
where
\begin{equation}
	\Delta^{-}C = - \frac{2}{N}\left[ \sum_{i\in{N_{pq}}}{\frac{1}{k_{i}(k_{i}-1)}} + \sum_{i\in\{p,q\}}{\left(\frac{|N_{pq}|}{(k_{i}-1)(k_{i}-2)} - \frac{C_{i}}{k_{i}-2}\right)}  \right].
\end{equation}

{\bf Assortativity Coefficient}

The updating formulas for $u, v, w$ are:
\begin{align}
	\widehat{u} &= u - 2 \left(\sum_{i\in N(p)}{k_{i}}+\sum_{i\in N(q)}{k_{i}} \right)-2(k_{p}-1)(k_{q}-1), \nonumber\\
	  \widehat{v} &= v - 4(k_{p}+k_{q}-1), \\
	  \widehat{w} &= w - 6\left[k_{p}(k_{p}-1)+k_{q}(k_{q}-1)\right] - 4 \nonumber.
\end{align}
Let
\begin{align}
	\Delta^{-}{u} &= - 2\left(\sum_{i\in N(p)}{k_{i}}+\sum_{i\in N(q)}{k_{i}}\right)-2(k_{p}-1)(k_{q}-1), \nonumber\\
	  \Delta^{-}{v} &= - 4\left(k_{p}+k_{q}-1 \right),\\
	  \Delta^{-}{w} &= - 6\left[k_{p}(k_{p}-1)+k_{q}(k_{q}-1)\right] - 4 \nonumber.
\end{align}
Then we have:
\begin{equation}
	\widehat{u} = u + \Delta^{-}{u}, \quad \widehat{v} = v + \Delta^{-}{v}, \quad \widehat{w} = w + \Delta^{-}{w},
\end{equation}
and the new assortativity coefficient $\widehat{r}$ is given by:
\begin{equation}
	\widehat{r} = \frac{8\widehat{M}\widehat{u}-\widehat{v}^2}{4\widehat{M}\widehat{w}-\widehat{v}^2}
	                     = \frac{8(M-1)(u+\Delta^{-}{u}) - (v+\Delta^{-}{v})^2}{4(M-1)(w+\Delta^{-}{w}) - (v+\Delta^{-}{v})^2}.
\end{equation}

{\bf Modularity}

For modularity, we again assume that the community partitions $g_{i}$ are unchanged after disconnecting the edge between $p$ and $q$. It follows that 
\begin{equation}
	\widehat{S}_A = S_A + \Delta^{-}S_A = S_A - 2 \delta(g_p,g_q), \quad ,
\end{equation}
\begin{equation}
	\widehat{S}_P = S_P + \Delta^{-}S_P = S_P - 2\left(K_{g_p} + K_{g_q}\right) + 2 \left[ \delta(g_p,g_q)+1\right]
\end{equation}
where $K_{g}$ is now updated using:
\begin{equation}
	\widehat{K}_g = K_g + \Delta^{-}K_g = K_g - \delta(g_p,g) - \delta(g_q,g).
\end{equation}
These now define the updating scheme for $\widehat{Q} = \left( \widehat{S}_A -\widehat{S}_P \slash2\widehat{M} \right) \slash2\widehat{M}$.

\subsection{On Computational Complexity}
In Table.~\ref{table:nonlin} we compare the computational complexity of using updating scheme (that depends on existing knowledge of statistics) and regular methods. Note that for regular methods the operations count depends on the data structure used to represent the network, and will be different in general. The updating scheme requires $O(1)$ operations to update for sparse graphs and at most $O(<k>)$, which has significant advantage comparing to regular method if graph size becomes large.

\begin{table}[ht]
\caption{Comparison of Computational Complexity}
\centering
\begin{tabular}{c c c c}
\hline\hline
Statistics & Updating Scheme & Adjacency Matrix & Edge List \\ [0.5ex]
\hline
degree (one node) 								& $O(1)$ 		& $O(N)$			& $O(<k>)$ \\
degree (network)  								& $O(1)$ 		& $O(N^2)$ 		& $O(<k>N)$ \\
clustering coefficient (one node) & $O(1)/O(<k>)$ 		& $O(<k>N)$ 	& $O(<k>^3)$ \\
clustering coefficient (network) 	& $O(<k>)$ 		& $O(<k>N^2)$ & $O(<k>^3N)$ \\
assortativity coefficient 				& $O(<k>)$ 	& $O(N^2)$		& $O(<k>N)$ \\
modularity measure 								& $O(1)$ 		& $O(N^2)$ 		& $O(<k>N)$ \\ [1ex]
\hline
\end{tabular}
\label{table:nonlin}
\end{table}

Our primary focus is developing efficient algorithms for application to problems of dynamic networks, and the computation savings is significant.  However, one may also consider the process of building a network, which can be viewed simply as an edge-adding algorithm from a starting set of a graph with $N$ nodes and no edges.  Then it takes $\frac{<k>N}{2}$  steps to create the network. 
So the formulas for degree and modularity indicate that computing the entire time sequence of statistics has the same computational complexity as doing the single computation for the final state (using the edge list). 
The formula for clustering coefficient is more efficient to calculate each value along the way rather that the single computation of the final state, although we also need to take the operations of building the network into account and (possibly) extra storage.
The time vector of assortativity coefficients requires an additional factor $<k>$, which is a minor price.

\section{Examples of Application}
In this subsection we show implementation of the above formula to obtain the evolution of some network statistics. We will focus on the case of adding edges between existing nodes, the other two operations will be very similear. The statistics we will calculate are the degree distribution, average clustering coefficient and modularity measure, although again, the evolution of other statistics can be obtained in the same manner by using the updating scheme. The evolving network models we choose are not intented to mimic real-world nets, but to show the efficiency of the updating scheme.

\subsection{Evolution of Degree and Clustering Coefficient}
We implement the updating scheme to track the evolution of degree distribution and average clustering coefficient of a growing random graph \cite{BOLLOBAS_BOOK}.

The growing graph is obtained as following:
start with a random graph of fixed size $N=1000$ with average degree $<k>=10$, then at each time step, randomly choose two nodes that are not connected, and make an edge between them, until the average degree of the network reaches $\widetilde{<k>}=20$. 

\begin{figure}[ht]
\includegraphics*[width=0.32\textwidth]{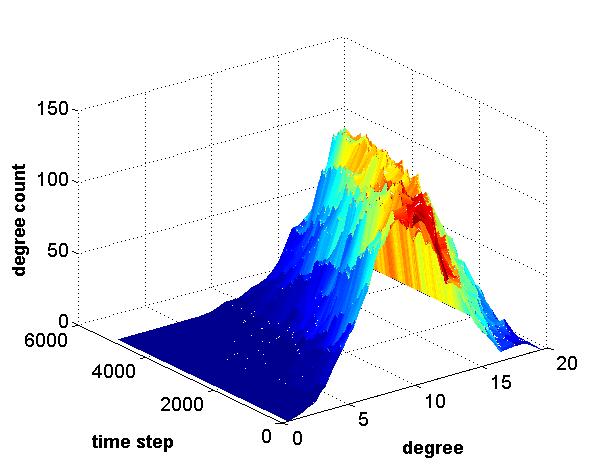}
\includegraphics*[width=0.32\textwidth]{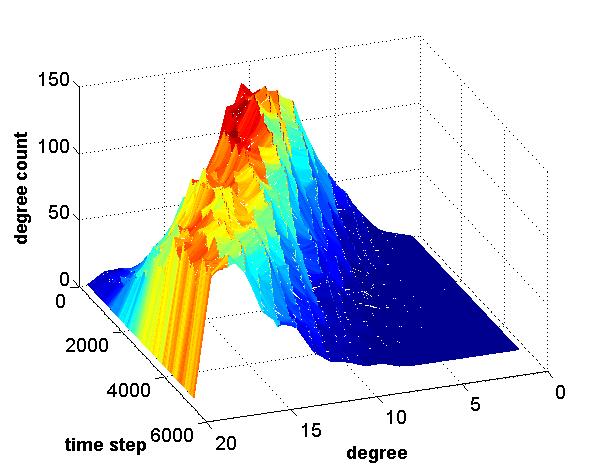}
\includegraphics*[width=0.32\textwidth]{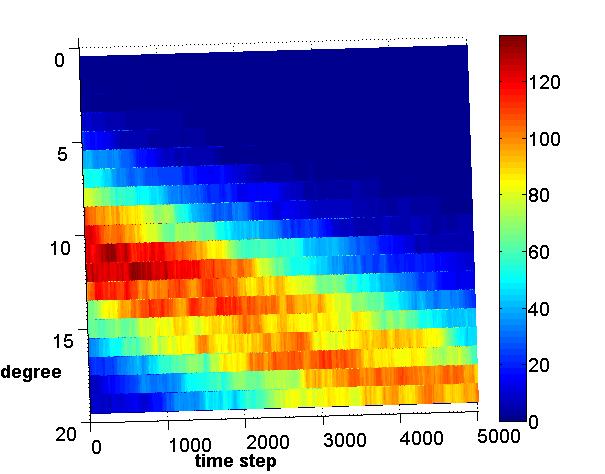}
\caption{(Color online) Evolution of degree distribution of a random growing network. The number of vertices is $5000$ in the network. Initially the connection probability of any pair of edge is $0.01$, by adding random edges in the network, this probability increases to $0.02$ in the end. We show three views of the evolution of the degree distribution as with respect to the process of add successive random edges. In the left and middle panel we see that for given time, the distribution is mimics a Possion distribution, and the peak is moving to larger degree side as time increases; while in the right panel we give a top view of the evolution.}
\label{ER_deg}
\end{figure}

\begin{figure}[ht]
\includegraphics*[width=0.55\textwidth]{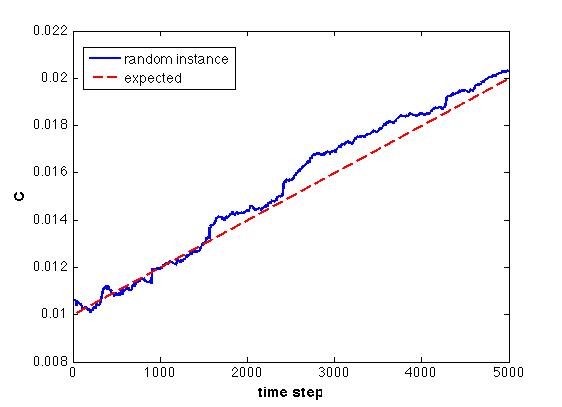}
\caption{(Color online) Evolution of the average clustering coefficient $C$ of a random growing network (described in Fig.~\ref{ER_deg}). Blue curve is the actual evolution of $C$, and red dashed line is the theoretical result given by $C_t=\frac{<k>}{N}$ where $<k>$ is the average degree at that time instant.}
\label{ER_clu}
\end{figure}

Fig.~\ref{ER_deg} and Fig.~\ref{ER_clu} shows the evolution of a typical realization of the above growing model. The total number of time steps is $5000$, which is $O(N)$ in this case. Note that using the updating scheme to obtain the evolution of degree in this case requires $O(N^2)$ (mostly for initial calculation) operations while using regular method would require $O(N^3)$ operations (using adjacency matrix); for average clustering coefficient the updating scheme requires $O(<k>N^2)$ operations and direct computation would require $O(<k>N^3)$ operations (also for adjacency matrix format). The above comparision holds very similarly for using edge list representation.

\subsection{Evolution of Modularity}
We artificially create an initial network with clear partition. The initial network is constructed as follows: generate an empty graph of $N$ vertices, prescribe a partition of the set $\{1,2,...,N\}$ into two groups such that the group sizes are $N_1,N_2$ (such that $N_1+N_2=N$) and probability $p_1,p_2,p_{between}$. Randomly connect any pair of vertices in group $1$ with probability $p_1$, and those in group $2$ with probability $p_2$; then randomly connect a vertice in group $1$ to a vertice in group $2$ with probability $p_{between}$. $p_{between}$ is usually chosen to be smaller than $p_1$ and $p_2$ so that the community structure is clear.

\begin{figure}[ht]
\includegraphics*[width=0.32\textwidth]{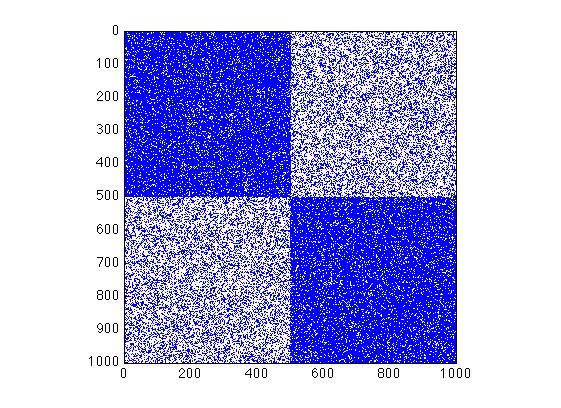}
\includegraphics*[width=0.32\textwidth]{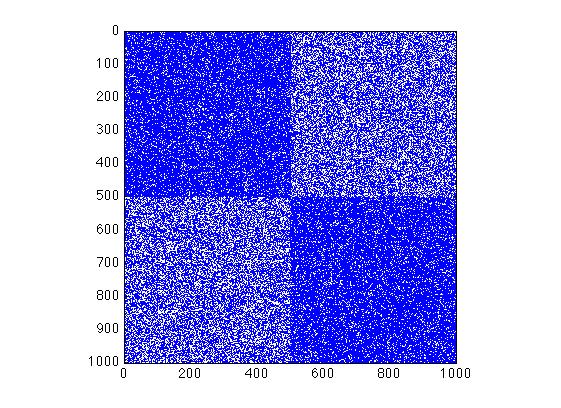}
\includegraphics*[width=0.32\textwidth]{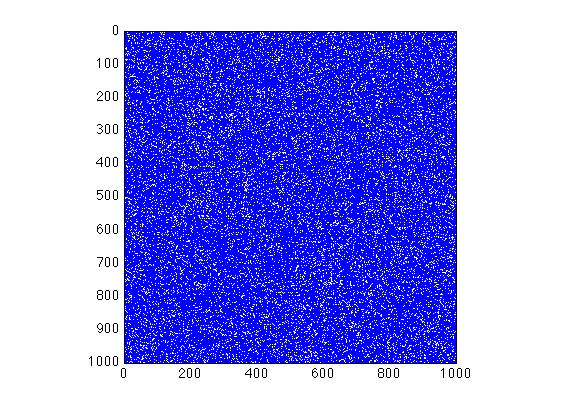}
\caption{Spy plot at three specific instances for the adjacency matrix. The left panel correponds to the initial network ($p_1=p_2=0.2$ and $p_{between}=0.05$), where there is a clear community structure. The middle panel corresponds to the time when $p_{between}$ reaches $0.1$ where the community structure becomes less apparent. The right panel is the end of the growing process such that $p_{between}=0.2$ and the network is totally random with no community structure.}
\label{Q_p_a}
\end{figure}

\begin{figure}[ht]
\includegraphics*[width=0.32\textwidth]{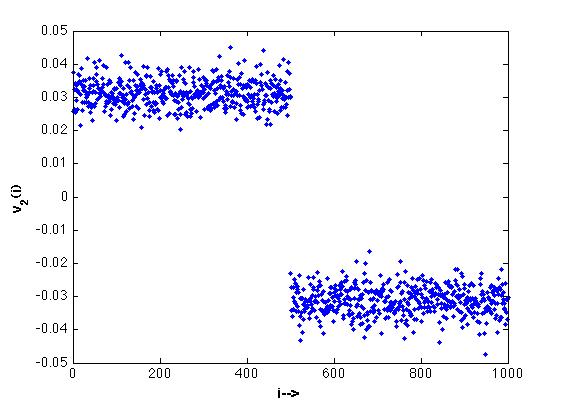}
\includegraphics*[width=0.32\textwidth]{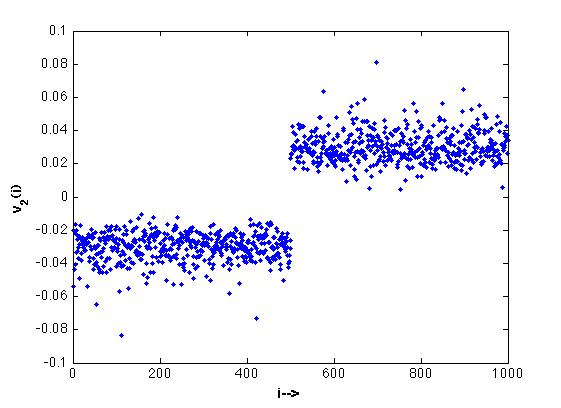}
\includegraphics*[width=0.32\textwidth]{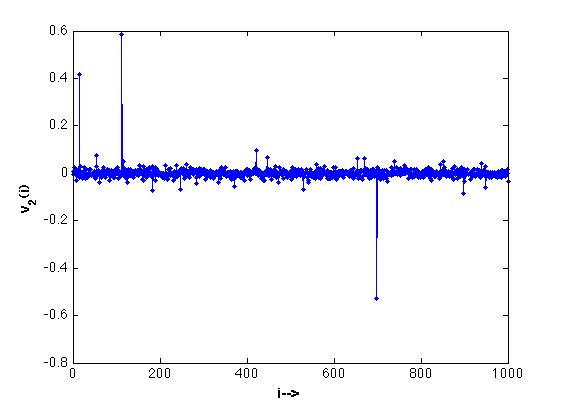}
\includegraphics*[width=0.32\textwidth]{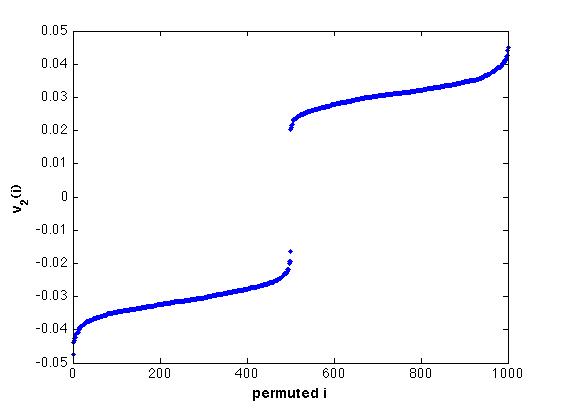}
\includegraphics*[width=0.32\textwidth]{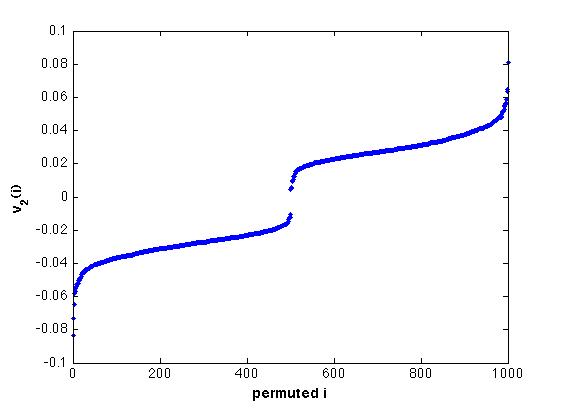}
\includegraphics*[width=0.32\textwidth]{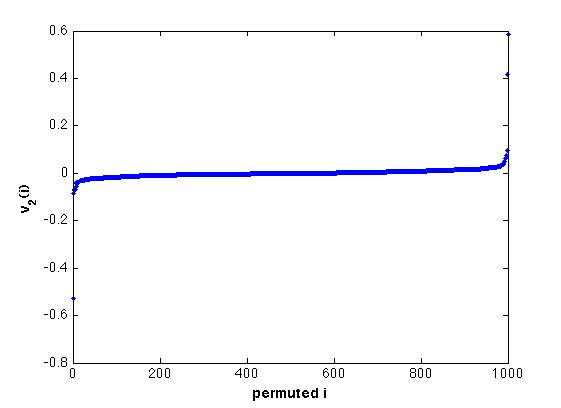}
\caption{Components of the Fiedler vector \cite{FIEDLER_89} at three specific time instances (see Fig.~\ref{Q_p_a}). In the three lower panels we plot the corresponding sorted components of the Fiedler vector.}
\label{Q_p_b}
\end{figure}

\begin{figure}[ht]
\includegraphics*[width=0.55\textwidth]{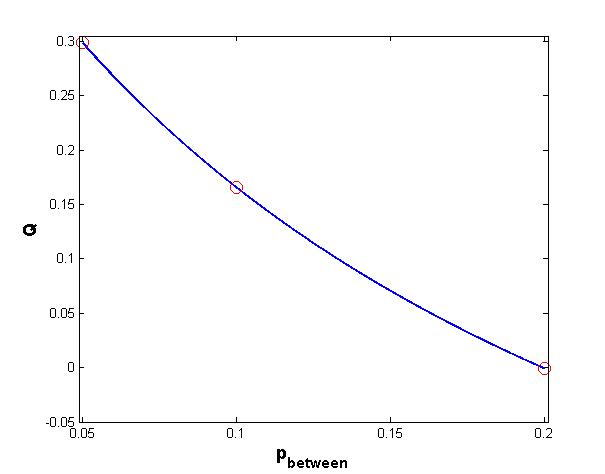}
\caption{(Color online) The evolution of modularity $Q$. Three red circles correspond to the time instances that are shown in Fig.~\ref{Q_p_a} and Fig.~\ref{Q_p_b}.}
\label{Q_evolution}
\end{figure}

In our example, we choose $N=1000$, group $1$ to be the set of nodes $\{1,2,...,500\}$ and the rest group $2$, so that $N_1 = N_2 = 500$. Also we let $p_1 = p_2 = 0.2$ and $p_{between}=0.05$. Then we add random edges between the groups until the probability of connecting in between groups are the same as the probability of connecting inside the groups (resulting in a completely random network in the end). Fig.~\ref{Q_p_a}, Fig.~\ref{Q_p_b} and Fig.~\ref{Q_evolution} shows the modularity affected by this process.

\section{Discussion and Conclusion}
In this paper we derive update formula for important network statistics (degree, clustering coefficient, assortativity coefficient, modularity), to provide theoretical tools for analyzing evolution of large evolving networks. The update formula are based on singe edge or node updating, while in general any updating of the graph can be decomposed into these basic one edge (node) operations and update using the formula we present in this paper. We also present several examples to illustrate the use of updating scheme, it is the use of update formula that allows us to efficiently track the evolution of network statistics, while traditional methods will require much more operations and become impratical.

The derivation of update formula requires that the statistics depend locally on network structure, for example, the update formula for clustering coefficient only requires the knowledge of local information of the vertices that are going to be connected. It becomes very hard, or maybe even impossible to derive exact update formula for statistics that depend on global information of the whole network, for example, the diameter, or the Fiedler vector of the network. However, the change of some of these global statistics can be bounded if there is only small change in the graph. For example, the change in the spectra and eigenvectors (including the Fiedler vector) of the graph Laplacian upon adding or deleting a few edges in the graph may be bounded by well-known perturbation results such as those in \cite{FIEDLER_89} and \cite{DEMMEL_BOOK}.

\acknowledgments
E.M.B, and S.J. are supported by the Army Research Office under 51950-MA. E.M.B. and J.D.S. are supported by the National Science Foundation under, DMS-0708083. J.P.B. gratefully acknowledges support from a National Science Foundation Graduate Research Fellowship.


\end{document}